# Cerebrovascular injury caused by a high strain rate insult in the thorax.

Amy C. Courtney, Ph.D., Force Protection Industries, Inc., 9801 Highway 78,
Ladson, SC, 29456 amy_courtney@post.harvard.edu

Michael W. Courtney, Ph.D., U.S. Air Force Academy, 2354 Fairchild Drive,
USAF Academy, CO, 80840-6210 Michael.Courtney@usafa.edu

Primary blast-induced traumatic brain injury (TBI) has increased in documented incidence and public prominence in recent conflicts. Evidence for a thoracic mechanism of blast-induced TBI was recently reviewed and, while the totality is compelling, data from experiments isolating this mechanism is sparse. Notably, one recent study showed pericapillar haemorrhage in brain tissue from victims of single, fatal gunshot wounds to the chest. Here, qualitative results are reported for a small field study that isolated a thoracic mechanism for TBI caused by a high strain rate insult in white-tailed deer (*Odocoileus virginianus*, mass 49-80 kg) in a natural environment. In each of three cases, petechiae were present on the surface of the frontal, occipital and/or left parietal lobes, along with capillary damage in the choroid plexus. The location of the projectile impact to the thorax seemed to affect the degree of damage. This may be due to the proximity to the great vessels. The data reported here provides direct evidence of a thoracic mechanism resulting in gross injury to the cerebral vasculature.

## Introduction

Primary blast-induced traumatic brain injury (TBI) is not new,[15] but it has increased in documented incidence (and public prominence) in modern conflicts.[13, 20] In order to better understand the mechanisms and pathology of TBI, laboratory experiments have been conducted that use shock tubes or actual blasts and animal models.[1, 4, 5, 14, 26] These studies have clarified the nature of the neural injury and suggested mechanical mechanisms by which the blast wave may result in brain injury.

At least three mechanical mechanisms for primary blast-induced TBI have emerged: direct cranial transmission of the blast wave, head acceleration, and a thoracic mechanism, the etiology of which includes a combination of kinetic energy transfer and a vagally-mediated response.[8] These mechanisms need not be mutually exclusive, and effective protection of soldiers may require improvements in more than one component of body armor. However, experimentally isolating a particular mechanism is challenging. Lacking direct experimental evidence, some have questioned whether a thoracic mechanism is relevant to TBI observed in recent conflicts.[21, 23]

It has been shown that a blast wave causes chest wall acceleration, which couples the wave into the thorax.[6] Similar insults, such as behind armor blunt trauma (BABT), in which a projectile strikes thoracic armor with enough energy to cause injury behind the armor, can also be quantified in terms of chest wall acceleration coupling a pressure wave into the underlying tissues.[8] BABT has been shown to cause remote cerebral effects in a large animal model.[11, 12] A penetrating bullet also transfers energy to fluids and tissues in the form of a pressure wave, and several studies have documented neural injury and remote cerebral effects by a penetrating projectile.[7, 18] Since this mechanism of neural injury does not involve other mechanisms, such as direct transmission of a pressure wave through the cranium or primary acceleration of the head, it can be used as a model to isolate a thoracic mechanism of TBI for study. Details about the application of physical principles to mathematically relate ballistic pressure wave insults to blast pressure wave insults have been published elsewhere.[8]

The characteristics of ballistic and blast pressure waves are similar. A ballistic pressure wave is generated when a ballistic projectile enters a viscous medium. A pressure wave in the thoracic cavity will refract through and reflect from internal structures, and the interaction of the resulting waves results in local pressure maxima. Tissue may be damaged anywhere the pressure magnitude is sufficiently large. Other characteristics of ballistic and blast pressure waves are also similar. Peak pressures are typically reached in a few microseconds, then pressure decreases exponentially in time over a pulse duration typically less than 2 ms. It is reasonable to

# Cerebrovascular injury caused by a high strain rate insult in the thorax

expect that similar waves would cause similar damage, and similar injuries have been reported, including cerebral effects.[8]

This paper presents qualitative results for a field study that isolated a thoracic mechanism for TBI caused by a high strain rate insult to the thorax in human-sized animals in natural conditions.

**Materials and Methods**

Three wild, white-tailed deer (*Odocoileus virginianus*, mass 49-80 kg) were shot broadside in the thorax with a single, rapidly expanding rifle bullet[a] (5.5 g, 6.53 mm diameter, impact velocity ~ 950 m/sec) transferring approximately 2500 J of energy.[b] The study design included antlerless subjects in a natural, unalarmed state and impacted in the thoracic cavity with no perforation of the diaphragm or superior thoracic structures.

Intact specimens were weighed and necropsy was immediately performed (less than 2 hours to completion), including dissection of the skull and removal of the intact brain and brainstem. Energy transfer characteristics and distance the deer traveled until incapacitation were recorded, and gross inspection of the thoracic walls, organs and brain was performed.

**Results**

In each case, no hematomas or large amounts of bleeding were observed in the brain. The bullet impacts were within the thoracic cavity, as specified by the study design, and within a few centimeters of each other. Impacts happened to occur in different locations relative to the heart and great vessels, and the extent of brain damage differed noticeably.

In the first case, there was no gross evidence of blood in the cerebrospinal fluid (CSF), and blood vessels appeared mostly intact on the cerebral cortex. However, petechiae were observed locally on the occipital lobe. In the second case, petechiae were noticeable on the surface of the frontal, occipital and left parietal lobes, along with capillary damage in the choroid plexus. In the third case, focal and diffuse vascular damage was observed in the subarachnoid space; the substance of the midbrain and the pituitary gland were pink with petechiae. Details of each case are summarized in Table 1.

---

[a] Winchester Ballistic Silvertip, model 51045
[b] Deer were harvested in accordance with the laws and rules of the Michigan Department of Natural Resources.

*Table 1* Summary of high strain rate experiment supporting a thoracic mechanism of TBI.

| | Mass (kg) | Brain Mass (g) | Impact Energy (J) | Distance to Incapacitation (m) |
|---|---|---|---|---|
| Antlerless Male | 49 | 179 | 2706 | 59 |
| Impact located 4 cm above midline, entered striking rib 7, exited between ribs 6 and 7. Capillary damage: petechiae observed on the occipital lobe, no hematomas. | | | | |
| Female | 80 | 170 | 2473 | 48 |
| Impact located 4 cm above midline, entered between ribs 8-9, grazed dorsal surface of liver, bullet recovered at 36 cm penetration. Capillary damage: petechiae observed on occipital, frontal and left parietal lobes and choroid plexus. | | | | |
| Female | 66 | 159 | 2445 | 16 |
| Impact located 12 cm below midline, entered between ribs 4-5, grazed the ventral surface of heart, exited breaking rib 3. Remarkably greater amount of vascular damage, midbrain and pituitary gland stained light red by diffuse petechiae. | | | | |

**Discussion**

Few experimental studies have attempted to isolate a thoracic mechanism of blast-induced TBI. Cernak et al. directed a shock tube at the thorax of a small animal model and observed damage to the brain.[4] However, possible confounding of a direct cranial mechanism cannot be ruled out due to diffraction of the shock wave at the opening of the shock tube. Saljo et al. also noted the possibility of this confounding factor in their experiment, in which pigs were exposed to blast waves on the abdomen or top of the head using a shock tube.[23] Romba et al. thoroughly shielded the head of a nonliving rhesus monkey instrumented in the thorax and brain with piezoelectric pressure sensors and detected a pressure impulse in the brain from blast exposure of the thorax.[22]

In the present study, vascular damage was observed in the brain that was visible to the unaided eye and that resulted from a remote ballistic impact to the thorax in human-sized animals. Knudsen and Oen observed dramatic remote cerebral effects in 8500 kg whales struck in the thorax or abdomen with grenade-tipped harpoons.[16] Using light microscopy, Suneson et al. observed blood-brain barrier damage in pigs that were shot in the thigh (distance to the brain was

# Cerebrovascular injury caused by a high strain rate insult in the thorax

approximately 0.5 m).[25] A recent review examined evidence for a thoracic mechanism from additional studies involving ballistic impacts and behind-armor blunt trauma.[8] Of particular relevance, using light microscopy, Krajsa observed cufflike pattern hemorrhages around small brain vessels in samples taken at autopsy from 33 human victims of single, fatal gunshot wounds to the chest (carefully selected to exclude patients with any related history).[17] He concluded:

> These haemorrhages are caused by sudden changes of the intravascular blood pressure as a result of a compression of intrathoracic great vessels by a shock wave caused by a penetrating bullet.

A recent study employed physical principles to relate ballistic pressure wave insults to blast pressure wave insults to the thorax.[9] That work introduced a region of increasing risk of blast-induced TBI due to a thoracic mechanism in terms of peak pressure and positive pulse duration. Using the methods described in that study, the impact energy in the current study was related to peak effective overpressure and positive pulse duration. The results are plotted in Figure 1 along with the region of increasing risk developed by Courtney and Courtney[9] and the Bowen curves for risk of blast-induced lung injury.[2]

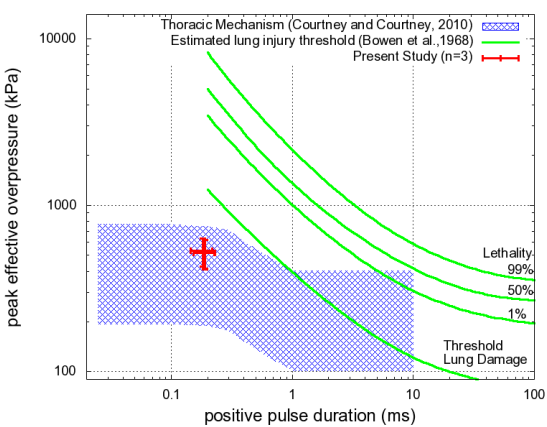

Figure 1. Experimental conditions for the present study were at the upper range of the region of increasing risk for a thoracic mechanism of blast-induced TBI using the methods described by Courtney and Courtney.[9]

In the present study, the location of bullet impact seemed to affect the degree of damage observed. Though the impact energy and distance to the brain in each case was similar, in the third case, the bullet penetrated closest to the heart and this animal was most quickly incapacitated. Also, the brain sustained visibly more damage in the third case compared to the first two. This result seems to support the specific hypothesis of vascular transmission of the pressure wave from the thorax to the brain. Subjects impacted further from the major vessels showed less grossly apparent brain injury and slower incapacitation. Alternatively, it could also be that the impact location increased pressures transmitted to the brain due to focusing effects of the reflected pressure waves from the thoracic walls.[19, 24]

## Conclusion

In the present study, the response of human-sized test subjects was investigated in a natural environment that minimized possible confounding of additional injury mechanisms. The origin of the pressure waves was confined to the thoracic cavity, so that no direct cranial transmission was possible. The bullet impact did not result in gross acceleration of the body as a whole or tertiary injury due to the head impacting the ground. Thus the study design isolated the thoracic mechanism. It is notable that cerebrovascular damage was readily observable in this study, and that the location of the bullet impact seemed to affect the degree of damage. These results support the idea that a thoracic mechanism of traumatic brain injury can result from expected levels of insult from some ballistic impacts and blast exposures.

## Acknowledgments

The authors are grateful for helpful comments from peer reviewers. This work was supported in part by BTG Research (www.btgresearch.org).

# Cerebrovascular injury caused by a high strain rate insult in the thorax